\documentclass[%
 reprint,
nofootinbib,
 amsmath,amssymb,
 aps,
floatfix,
]{revtex4-1}

\usepackage{graphicx}
\usepackage{dcolumn}
\usepackage{bm}
\usepackage{amsmath}
\usepackage{amssymb}
\usepackage{hyperref}
\usepackage[capitalise]{cleveref}
\usepackage[dvipsnames]{xcolor}

\crefname{chapter}{chap.}{chap.}
\crefname{section}{Sec.}{Sec.}
\Crefname{chapter}{Chapter}{Chapters}
\Crefname{section}{Section}{Sections}

\begin{document}


\title{Constraints on the cosmic string loop collapse fraction from Primordial Black Holes}
\author{Chloe James-Turner}
\affiliation{School of Physics and Astronomy, University of Nottingham, University Park, Nottingham, NG7 2RD, United Kingdom}
\author{Danton P. B. Weil}
\affiliation{School of Physics and Astronomy, University of Nottingham, University Park, Nottingham, NG7 2RD, United Kingdom}
\author{Anne M. Green}
\email{anne.green@nottingham.ac.uk}
\affiliation{School of Physics and Astronomy, University of Nottingham, University Park, Nottingham, NG7 2RD, United Kingdom}
\author{Edmund J. Copeland} 
\affiliation{School of Physics and Astronomy, University of Nottingham, University Park, Nottingham, NG7 2RD, United Kingdom}

\date{\today}

\begin{abstract}
A small fraction, $f$, of cosmic string loops can collapse to form Primordial Black Holes (PBHs). Constraints on the abundance of PBHs can therefore be used to constrain $f$. We update these calculations, taking into account the PBH extended mass function, and find 
$f < 10^{-31} (G \mu/ c^2)^{-3/2}$.
This is roughly one order of magnitude tighter than previous constraints. The improvement from the tighter constraints on the abundance of PBHs is partly off-set by refinements to the theoretical calculation of the cosmic string loop formation rate. 
\end{abstract}

\maketitle

\section{Introduction}

Cosmic strings are topological defects which may form during phase transitions in the early Universe~\cite{Kibble:1976sj}. For reviews see e.g.~Refs.~\cite{Vilenkin:2000jqa,Copeland:2009ga,Sakellariadou:2009ev,Copeland:2011dx}. The cosmic string network quickly reaches a stable scaling solution, with loops of radii smaller than the Hubble radius constantly being produced due to the dynamics of long string interactions and loop self-intercommutations. These loops oscillate and decay slowly by emitting gravitational radiation~\cite{Vachaspati:1984gt}.  A cosmic string loop which contracts under its own tension to a size smaller than its Schwarzschild radius can form a black hole~\cite{Hawking:1987bn,Polnarev:1988dh}. There are tight observational constraints on the abundance of Primordial Black Holes (PBHs), in particular from the products of their evaporation~\cite{Carr:2009jm}. These constraints can be translated into limits on the cosmic string loop collapse fraction~\cite{Caldwell:1993kv,MacGibbon:1997pu}.

The cosmic string tension, $(G \mu/c^2)$, plays a key role in the properties of the cosmic string network. The tightest current constraint on its value comes from limits, from the Parkes Pulsar Timing Array~\cite{Shannon:2015ect,Lasky:2015lej}, on the stochastic gravitational wave background produced by oscillations of cosmic string loops. For the standard cosmic string production scenario, and assuming a string intercommutation probability $p=1$ as found in field theory simulations of 
local cosmic strings~\cite{Matzner}, $(G \mu/c^2) < 1.5 \times 10^{-11}$~\cite{Blanco-Pillado:2017rnf}. 

As we will outline in Sec.~\ref{sec:sc}, the loop formation rate, and hence the PBH formation rate, ${\rm d} n_{\rm PBH}/ {\rm d} t$, is proportional to $t^{-4} a^3(t)$ where $a(t)$ is the scale factor. During radiation domination, $a(t) \propto t^{1/2}$, while the mass, $M$, of a PBH formed from cosmic string loop collapse at time $t$ is proportional to the typical string loop mass, which varies as $t$. Consequently the PBHs have a mass function of the form ${\rm d} n_{\rm PBH}/ {\rm d} M \propto M^{-5/2}$~\cite{MacGibbon:1997pu}.
This extended mass function needs to be taken into account when applying the PBH abundance constraints~\cite{MacGibbon:1997pu}, which are usually calculated assuming a delta-function mass function.

In Sec.~\ref{sec:sc} we outline the calculation of the cosmic string loop formation rate and the resulting PBH abundance. In Sec.~\ref{sec:pbh} we use updated PBH abundance constraints to constrain the properties of the cosmic string network, in particular the loop collapse fraction, $f$. Finally we give concluding remarks in Sec.~\ref{sec:dis}.

\section{Cosmic string loop formation and collapse}
\label{sec:sc}

The density parameter of PBHs produced by cosmic string loops was previously derived by Caldwell \& Gates~\cite{Caldwell:1993kv} and MacGibbon et al.~\cite{MacGibbon:1997pu}, using the calculation of the cosmic string loop production rate from Caldwell \& Allen~\cite{Caldwell:1991jj}. We outline the calculation here in order to clarify numerical factors in the final results~\footnote{This calculation could alternatively, in principle, be carried out in terms of the loop production rate found from simulations e.g. Ref.~\cite{Blanco-Pillado:2013qja}. }. 
Throughout we assume the standard scenario, where cosmic strings are produced in a thermal phase transition and reach a scaling solution soon afterwards. 

Long cosmic strings, denoted by `$\infty$', have equation of state
\begin{equation}
  p_{\infty}=\frac{1}{3}\rho_{\infty} c^2 [2\langle v^2\rangle-1] \,,
    \label{eqn:EoScosmicstrings}
\end{equation}
where $\langle v^2 \rangle$ is the mean squared velocity in units of $c^2$, and recent simulations find that during radiation domination 
$\langle v^2 \rangle = 0.40$~\cite{BlancoPillado:2011dq}.

Cosmic string loops, denoted by `$\ell$', have $p_{\ell}=0$ and hence, using conservation of energy in an expanding universe,
\begin{equation}
    \dot{\rho}_{\infty}+2H\rho_{\infty}[\langle v^2\rangle+1]=
    -[\dot{\rho}_{\ell}+3H\rho_{\ell}] \,,
    \label{eqn:fluid2}
\end{equation}
where dots denote differentiation with respect to time and $H=\dot{a}/a$ is the Hubble parameter.
The right hand side of Eq.~(\ref{eqn:fluid2}) corresponds to the rate of change in energy due to loops being formed. This energy can be related to the number of loops produced up to time $t$ within a comoving volume $V(t)$, $n_{\ell}$:
\begin{equation}
    \frac{{\rm d} E_{\ell}}{dt}= \mu c^2 L \frac{{\rm d}n_{\ell}}{{\rm d} t}, 
    \label{eqn:CSLdE/dt}
\end{equation}
and $ E_{\ell} = V(t) \rho_{\ell}$.

Following Refs.~\cite{Caldwell:1993kv,MacGibbon:1997pu}, and also more recent work constraining cosmic strings using LIGO data~\cite{Abbott:2017mem}, we assume that all of the loops formed have a length which is a constant fraction of the horizon distance at that time:  $L=\alpha R_{\rm H} (t)$ where 
\begin{equation}
R_{\rm H}(t)= c a(t) \int_{0}^{t} \frac{{\rm d} t^{\prime}}{a(t^{\prime})} \,,
\end{equation}
is the horizon distance.  Recent large simulations find $\alpha \sim 0.1$~\cite{BlancoPillado:2011dq}.
Refs.~\cite{Caldwell:1993kv,MacGibbon:1997pu} assumed that all of the loop energy is in these large loops. However more recent simulations (e.g. Ref.~\cite{Blanco-Pillado:2013qja}) have subsequently found that only a small fraction $\delta \sim 0.1$ of the loop energy is in large loops, with the majority going into the kinetic energy of the subdominant (in terms of number density) small loop population, which red-shifts away quickly. The infinite string density is given by~\cite{Caldwell:1991jj}
\begin{equation}
\label{rholong}
   \rho_{\infty} = \frac{A\mu c^2}{R_{\rm H}^2(t)} \,,
\end{equation}
where recent simulations find $A \approx 44$~\cite{BlancoPillado:2011dq,Blanco-Pillado:2019vcs}.

Combining Eqs.~(\ref{eqn:fluid2}), (\ref{eqn:CSLdE/dt}) and (\ref{rholong}), and taking into account the fraction of the loop energy in long loops, $\delta$, 
the rate of loop formation during radiation domination is given by
\begin{equation}
    \frac{{\rm d} n_{\ell}}{{\rm d} t}=
    \frac{A \delta  \; V(t)}{8  \alpha c^{3}t^{4} }[1- \langle v^2 \rangle] \,,
    \label{eqn:looprate3}
\end{equation}
and PBHs form from the collapse of cosmic string loops at a rate 
\begin{equation}
    \frac{{\rm d} n_{\rm PBH}}{{\rm d} t} = f \frac{{\rm d} n_{\ell}}{{\rm d} t} \,.
    \label{eqn:fraction1}
\end{equation}
The loop collapse fraction, $f$, is not well known, in particular for the observationally allowed values of $(G \mu /c^2)$. 
Following earlier analytic calculations~\cite{Hawking:1987bn,Polnarev:1988dh}, numerical simulations by Caldwell \& Casper found that $f = 10^{4.9 \pm 0.2} (G \mu/c^2)^{4.1 \pm 0.1}$ for $10^{-3} \lesssim (G \mu/c^2) \lesssim 3 \times 10^{-2}$~\cite{Caldwell:1995fu}. They argued that this parameterization is valid down to $( G \mu /c^2) \sim10^{-6}$, however the observational constraints from the stochastic gravitational wave background are now much tighter than this, $( G \mu /c^2) \lesssim10^{-11}$.
We therefore treat $f$ as an unknown parameter.

 In addition to the factor of $\delta$, our expression for the loop collapse rate, ${\rm d} n_{\ell} / {\rm d} t$, Eq.~(\ref{eqn:looprate3}), differs from that used by MacGibbon et al.~\cite{MacGibbon:1997pu} by a factor $([1-\langle v^2 \rangle] /8 )/4 \sim 10^{-2}$ which is non-negligible. They cite Caldwell \& Gates~\cite{Caldwell:1993kv} for the origin of their expression, who in turn cite Caldwell \& Allen~\cite{Caldwell:1991jj}. Caldwell \& Allen's final expression for the loop collapse rate, their Eq.~(3.7), is written in the form
\begin{equation}
 \frac{{\rm d} n_{\ell}}{{\rm d} t}=
  -  \frac{ V(t)}{\mu \alpha c^{2} R_{\rm H}(t)}   \left[ \dot{\rho}_{\infty}+2H \rho_{\infty} \left( \langle v^2\rangle+1 \right) \right] \,.
\end{equation}
Our expression, Eq.~(\ref{eqn:looprate3}), comes from substituting their expression for $\rho_{\infty}$, Eq.~(\ref{rholong}), into this expression along with $H=1/(2t)$ and $R_{\rm H}(t) = 2 c t$. The factor of $8$ in our expression comes from three powers of the factors of 2  in the Hubble parameter and the horizon distance. This suggests that Ref.~\cite{Caldwell:1993kv} may have made the approximations $H \approx 1/t$ and $R_{\rm H}(t) \approx ct$. The additional factor of $4$ is in fact absorbed by the different fiducial values of $A$, the constant of proportionality in the expression for $\rho_{\infty}$. As discussed above, we use $A \approx 44$ which is similar to the value found by Caldwell \& Allen~\cite{Caldwell:1991jj} for radiation domination ($A\approx 52$), while Caldwell \& Gates~\cite{Caldwell:1991jj} and MacGibbon et al.~\cite{MacGibbon:1997pu} use $A \approx 10$. So ultimately our expression for the loop formation rate (and hence the PBH formation rate) is smaller by a factor $(1-\langle v^2 \rangle ) /8 =0.075$.

The present day PBH density parameter is given by
\begin{equation}
    \Omega_{\rm{PBH}}= \frac{1}{\rho_{\rm{c},0}} \int_{t_{\star}}^{t_0}   \frac{{\rm d} n_{\rm{PBH}}}{{\rm d} t^{\prime}} M(t^{\prime}) {\rm d} t^{\prime} \,,
    \label{eqn:omega0}
\end{equation}
where $\rho_{{\rm c},0}$ is the critical density today, at $t_{0}$, and $M(t^{\prime}) \approx \alpha \mu c t^{\prime}$ is the mass of a PBH that forms at time $t^{\prime}$~\footnote{The PBH mass may depend on the radius of the loop from which it formed~\cite{Helfer:2018qgv}, however this has to date only been studied for large tensions, $ (G \mu/c^2 )\sim 10^{-2}$, therefore we do not take this into account. We also do not include the decrease in the mass of an individual PBH with time due to its evaporation, as this is only significant at the very end of its lifetime, e.g. Ref.~\cite{Carr:2016hva}.}
and $t_{\star}$ is the formation time of a PBH with mass $M_{\star} \approx 5 \times 10^{14} \, {\rm g}$ which is completing its evaporation today~\cite{MacGibbon:2007yq}. We assume that the mass of the lightest PBHs formed, $M_{\rm min} = \alpha \mu c t_{\rm min}$ where $t_{\rm min}$ is the formation time of the first lightest PBHs, is much smaller than $M_{\star}$. Using the relationship between temperature and time during radiation domination~\cite{Kolb:1990vq}
\[
\left( \frac{t}{1 \, {\rm s}} \right) \approx \left( \frac{1 \, {\rm MeV}}{T} \right)^{2} \,,
\]
this is the case if the phase transition at which the cosmic strings form takes place at a temperature
\[
 T \gg \left(\frac{G \mu/c^2}{10^{-11}} \right)^{1/2} \, 10^{3} \, {\rm GeV} \,.
 \]

It will prove useful to define an effective present day PBH density parameter, which is the density the PBHs would have today if those with $M< M_{\star}$ had not evaporated:
\begin{equation}
\label{omegaeff1}
     \Omega_{\rm{PBH}}^{\rm eff}= \frac{1}{\rho_{\rm{c},0}} \int_{t_{\rm min}}^{t_0}   \frac{{\rm d} n_{\rm{PBH}}}{{\rm d} t^{\prime}} M(t^{\prime}) {\rm d} t^{\prime} \,.
\end{equation}
Inserting Eqs.~(\ref{eqn:looprate3}) and (\ref{eqn:fraction1})  into Eq.~(\ref{omegaeff1}), and neglecting the recent period of dark energy domination so that $\rho_{{\rm c},0} = (6 \pi G t_{0}^2)^{-1}$ and
 \begin{equation}
V(t^{\prime}) \equiv \left[ \frac{a(t^{\prime})}{a(t_{0})} \right]^3= \left( \frac{t_{\rm eq}}{t_{0}} \right)^{2} 
   \left( \frac{t^{\prime}}{t_{\rm eq}} \right)^{3/2} \,,
\end{equation}
where $t_{\rm eq}$ is the time at matter-radiation equality, we find
\begin{equation}
    \Omega_{\rm{PBH}}^{\rm{eff}} = \frac{3\pi}{2}
     A  \delta f  [1- \langle v^{2}\rangle]  \frac{G \mu}{c^2}  \left( \frac{t_{\rm min}}{t_{\rm eq}} \right)^{-1/2} \,.
    \label{eqn:omegaeff5}
\end{equation}
Using $t_{\rm eq} =  3.4 \times 10^{10}  (\Omega_{\rm m} h^2)^{-2} \, {\rm s} $~\cite{Kolb:1990vq} with $\Omega_{\rm m} h^2 =0.14$~\cite{Aghanim:2018eyx} this can be rewritten as
\begin{equation}
\label{omegaeff}
\Omega_{\rm PBH}^{\rm eff} (M_{\rm min}) \approx  c_{\rm string} \left( \frac{M_{\rm min}}{M_{\star}} \right)^{-1/2} \,,
\end{equation}
with 
\begin{eqnarray}
\label{cstring}
c_{\rm string} &= &3 \times 10^{2 }  f \left( \frac{ \delta}{0.1} \right)  \left( \frac{A}{44} \right)  \left( \frac{\alpha}{0.1} \right) ^{1/2} \left( \frac{ [1- \langle v^{2}\rangle] }{0.6} \right) 
\nonumber \\
 & \times &
   \left( \frac{G \mu/c^2}{ 1.5 \times 10^{-11}} \right)^{3/2} \,,
\end{eqnarray}
where we have normalized the string network parameters to the values discussed above.
Our expression for $\Omega_{\rm PBH}^{\rm eff}$ differs from previous calculations~\cite{MacGibbon:1997pu}, due to the
difference in the normalization of the cosmic string loop production rate discussed above and also the
inclusion of PBHs with $M< M_{\star}$. The latter allows us to include the tight constraints on the abundance of PBH with $M \sim 10^{13-14} \, {\rm g}$ 
which come from the damping of small-scale CMB anisotropies.

\section{Primordial Black Hole constraints}
\label{sec:pbh}

Constraints on the abundance of PBHs are usually calculated assuming a delta-function mass function, i.e. assuming that all PBHs form at the same time with mass equal to a constant fraction, $\gamma$, of the horizon mass at that time $M= \gamma M_{\rm H}$.
For PBHs with $M<M_{\star}$ which have evaporated by today the constraints are expressed in terms of the initial PBH density fraction
\begin{equation}
\beta(M) = \frac{\rho_{\rm PBH}(t_{\rm i})}{\rho_{\rm c}(t_{\rm i})} \,,
\end{equation}
where $t_{\rm i}$ is the formation time and $\rho_{\rm c}(t_{\rm i})$ is the critical density at that time.
Ref.~\cite{Carr:2009jm} presents constraints using $\beta^{\prime}(M)$ defined, so as to absorb the uncertainties which arise when evolving the constraint back to the PBH formation time, as  
\begin{equation}
\beta^{\prime}(M) \equiv \gamma^{1/2} \left( \frac{g_{\star, i}}{106.75} \right)^{-1/4} \beta(M) \,,
\end{equation}
where $g_{\star, i}$ is the number of relativistic degrees of freedom at the time the PBHs form. 
Assuming all PBHs form at the same time with a mass which is a constant fraction of the horizon mass, then the initial mass fraction is related to the present PBH abundance by~\cite{Carr:2009jm}
\begin{equation}
\label{omegabeta}
\Omega_{\rm PBH}^{\rm DF}(M) = \left( \frac{\beta^{\prime} (M)}{1.2 \times 10^{-8}} \right) \left( \frac{ h}{0.72} \right)^{-2}  \left( \frac{M}{M_{\odot}} \right)^{-1/2} \,,
\end{equation}
where $h$ is the dimensionless Hubble constant and `${\rm DF}$' emphasises that this expression is only valid if all PBHs form at the same time with a delta-function mass function.

Ref.~\cite{Caldwell:1993kv} found that the tightest constraints on the cosmic string loop collapse fraction $f$ came from the tightest constraints on the initial PBH mass fraction $\beta(M)$. We therefore concentrate on the tightest constraints on $\beta(M)$, which come from the effects of PBH evaporation at the present day, or in the (relatively) recent past.

For PBHs with mass $M< M_{\star}$ which have evaporated by today, secondary gamma-rays dominate the spectrum of gamma-rays produced and
\begin{equation}
    \beta^{\prime} (M) \lesssim 3 \times 10^{-27}  \left( \frac{M}{M_{*}} \right) ^{-5/2 - 2\epsilon} \hspace{0.5cm}  {\rm for} \,\, M<M_{*} \,,
    \label{eqn:gammacons1}
\end{equation}
where $\epsilon \sim 0.1-0.4$ parameterizes the energy dependence of the observed gamma-ray intensity: $I^{\rm obs} \propto E_{\gamma}^{-(1+\epsilon)}$\cite{Carr:2009jm}.

For PBHs which have not yet completed their evaporation primary gamma-rays dominate and~\cite{Carr:2009jm}
\begin{equation}
    \beta^{\prime}(M) \lesssim 4 \times 10^{-26}  \left( \frac{M}{M_{*}} \right) ^{7/2 + \epsilon} \hspace{0.5cm}  {\rm for} \,\,  M>M_{*} \,.
    \label{eqn:gammacons2}
\end{equation}
As pointed out by Ref.~\cite{Barrau:2003nj}, a tighter constraint can be found by taking into account the contribution of known astrophysical sources, such as blazars. This has recently been done in Ref.~\cite{Ballesteros:2019exr} and we also consider their constraint on $\Omega_{\rm PBH}^{\rm DF}(M)$. 

PBHs evaporating after recombination would damp small-scale CMB anisotropies leading to a constraint~\cite{Carr:2009jm}
\begin{eqnarray}
    \beta^{\prime}(M) &\leq& 3 \times 10^{-30}  \left( \frac{f_{\mathrm{H}}}{0.1} \right)^{-1} \left( \frac{M}{10^{13}\mathrm{g}} \right) ^{3.1} \,,
    \nonumber \\ 
 &&    {\rm for} \,\,  2.5 \times 10^{13} \, {\rm g} < M < 2.4 \times 10^{14} \, {\rm g} \,,
    \label{eqn:CMBcons}
\end{eqnarray}
where $f_{\mathrm{H}} \approx 0.1$ is the fraction of the emission in the form of electrons and positrons.

Constraints calculated assuming a delta-function mass function can be translated to an extended mass function using~\cite{Carr:2017jsz}~\footnote{We use slightly different notation to Ref.~\cite{Carr:2017jsz} as they were interested in PBHs as a dark matter candidate and hence only considered masses $M> M_{\star}$.}
\begin{equation}
\label{emf}
\int_{M_{1}}^{M_{2}} \frac{\psi(M)}{\Omega_{\rm PBH}^{\rm DF}(M)} \, {\rm d} M \leq 1 \,,
\end{equation}
where $M_{1} < M < M_{2}$ is the range of validity of the constraint and $\psi(M)$ is defined such that the present day effective PBH density parameter is given by
\begin{equation}
\Omega_{{\rm PBH}}^{\rm eff}  = \int_{0}^{\infty} \psi(M) \, {\rm d} M \,.
\end{equation} 
Using Eq.~(\ref{omegaeff}) we have
\begin{equation}
 \label{psi}
\psi(M) = \frac{c_{\rm string}}{2 M_{\star}} \left( \frac{M}{M_{\star} }\right)^{-3/2} \hspace{0.5cm} M_{\rm min} < M < M_{\rm max} \,, 
\end{equation}
where $M_{\rm min}$ and $M_{\rm max}$ are the lightest and heaviest PBHs formed and the dimensionless constant $c_{\rm string}$ depends on the parameters of the cosmic string network and is given by Eq.~(\ref{cstring}).

We use Eq.~(\ref{omegabeta}) to convert the constraints on $\beta^{\prime}(M)$, Eqs.~(\ref{eqn:gammacons1}-\ref{eqn:CMBcons}), into constraints on $\Omega_{\rm PBH}^{\rm eff}(M_{\rm min})$. We then use Eq.~(\ref{emf}) and the mass function of PBHs formed via cosmic string collapse, Eq.~(\ref{psi}), to covert each of the constraints into a constraint on $c_{\rm string}$ which parameterises the amplitude of the PBH mass function. Eq.~(\ref{cstring}) can then be used to translate this constraint into a limit on the cosmic string loop collapse fraction, $f$. In each case if $M_{\rm min} < M_{1}$, as expected as discussed above, then $\Omega_{\rm PBH}^{\rm eff}(M_{\rm min}) \propto M_{\rm min}^{-1/2}$ and the constraint on $c_{\rm string}$ is independent of $M_{\rm min}$:  
\begin{eqnarray}
\label{clim}
c_{\rm string} &< & \begin{cases}
2 \times 10^{-9}  & \text{primary gamma-rays  \,,} \\
1 \times 10^{-10}  & \text{secondary gamma-rays \,,} \\
 2 \times 10^{-12}  & \text{CMB anisotropies \,.} \\
\end{cases}
\end{eqnarray}

Using Eq.~(\ref{cstring}), the tightest constraint on $c_{\rm string}$ in Eq.~(\ref{clim}), from the damping of small-scale CMB anisotropies, translates into a limit on the cosmic string loop collapse fraction, $f$:
\begin{eqnarray}
f &\lesssim & 7 \times 10^{-15 }  \left( \frac{\delta}{0.1} \right)^{-1}  \left( \frac{A}{44} \right)^{-1}  \left( \frac{\alpha}{0.1} \right) ^{-1/2}
 \nonumber \\ & \times &
 \left( \frac{ [1- \langle v^{2}\rangle] }{0.6} \right)^{-1} \left( \frac{G \mu/c^2}{ 1.5 \times 10^{-11}} \right)^{-3/2} \,.
\end{eqnarray}
The constraints from secondary and primary gamma-rays lead to limits on $f$ that are two and three orders of magnitude weaker respectively. 

\section{Summary}
\label{sec:dis}

We have updated the constraints on the collapse of cosmic string loops from limits on the abundance of PBHs. 
In Sec.~\ref{sec:sc} we revisited the calculation of the rate at which cosmic string loops form and the resulting PBH mass function~\cite{Caldwell:1991jj,MacGibbon:1997pu}. We first of all incorporated the impact of recent numerical simulations \cite{Blanco-Pillado:2013qja} which indicate that only a fraction $\delta \sim 0.1$ of the energy of the long strings go into the numerically dominant large loops. Secondly we found a loop (and hence PBH) formation rate which is smaller than in Refs.~\cite{Caldwell:1991jj,MacGibbon:1997pu} by a factor of $0.075$. We believe that this difference comes from the inclusion of numerical factors in the expressions for the horizon distance and Hubble parameter.

In Sec.~\ref{sec:pbh} we then applied the tightest current limits on the abundance of PBHs, from extra-galactic gamma-rays and the damping of small-scale CMB anisotropies~\cite{Carr:2009jm}, to PBHs formed from cosmic string collapse. These limits were calculated assuming a delta-function PBH mass function. We therefore used the formalism of Ref.~\cite{Carr:2017jsz} to apply them to the $M^{-5/2}$ mass function of PBHs formed from cosmic string collapse. The tightest constraint on the amplitude of the PBH mass function comes from the damping of small-scale CMB anisotropies. Using values for the cosmic string network parameters from recent simulations~\cite{BlancoPillado:2011dq}, it leads to a limit on the cosmic string loop collapse rate: $f \lesssim 7 \times 10^{-15 }  [(G \mu/c^2)/ 10^{-11}]^{-3/2}$. Here we have normalised the limit to the tightest constraint on the cosmic string tension, from limits on the stochastic gravitational wave background produced by oscillations of cosmic string loops~\cite{Blanco-Pillado:2017rnf}. 

Our limit is roughly one order of magnitude tighter than that found in Ref.~\cite{MacGibbon:1997pu} using gamma-ray constraints on the PBH abundance. The CMB constraint we use is much tighter than this gamma-ray constraint. However the resulting improvement in the cosmic string loop collapse rate is partly off-set by our refinement of the theoretical calculation of the cosmic string loop formation rate. 

\section*{Acknowledgements}
We would like to thank Robert Caldwell and Bernard Carr for helpful discussions, and one of the referees for pointing out that only a small fraction of  the energy density in the string network goes into large loops. EJC and AMG acknowledge support from STFC grant ST/P000703/1.

\section*{References}

\bibliography{mybibfile}

\begin{thebibliography}{29}%
\makeatletter
\providecommand \@ifxundefined [1]{%
 \@ifx{#1\undefined}
}%
\providecommand \@ifnum [1]{%
 \ifnum #1\expandafter \@firstoftwo
 \else \expandafter \@secondoftwo
 \fi
}%
\providecommand \@ifx [1]{%
 \ifx #1\expandafter \@firstoftwo
 \else \expandafter \@secondoftwo
 \fi
}%
\providecommand \natexlab [1]{#1}%
\providecommand \enquote  [1]{``#1''}%
\providecommand \bibnamefont  [1]{#1}%
\providecommand \bibfnamefont [1]{#1}%
\providecommand \citenamefont [1]{#1}%
\providecommand \href@noop [0]{\@secondoftwo}%
\providecommand \href [0]{\begingroup \@sanitize@url \@href}%
\providecommand \@href[1]{\@@startlink{#1}\@@href}%
\providecommand \@@href[1]{\endgroup#1\@@endlink}%
\providecommand \@sanitize@url [0]{\catcode `\\12\catcode `\$12\catcode
  `\&12\catcode `\#12\catcode `\^12\catcode `\_12\catcode `\%12\relax}%
\providecommand \@@startlink[1]{}%
\providecommand \@@endlink[0]{}%
\providecommand \url  [0]{\begingroup\@sanitize@url \@url }%
\providecommand \@url [1]{\endgroup\@href {#1}{\urlprefix }}%
\providecommand \urlprefix  [0]{URL }%
\providecommand \Eprint [0]{\href }%
\providecommand \doibase [0]{http://dx.doi.org/}%
\providecommand \selectlanguage [0]{\@gobble}%
\providecommand \bibinfo  [0]{\@secondoftwo}%
\providecommand \bibfield  [0]{\@secondoftwo}%
\providecommand \translation [1]{[#1]}%
\providecommand \BibitemOpen [0]{}%
\providecommand \bibitemStop [0]{}%
\providecommand \bibitemNoStop [0]{.\EOS\space}%
\providecommand \EOS [0]{\spacefactor3000\relax}%
\providecommand \BibitemShut  [1]{\csname bibitem#1\endcsname}%
\let\auto@bib@innerbib\@empty
\bibitem [{\citenamefont {Kibble}(1976)}]{Kibble:1976sj}%
  \BibitemOpen
  \bibfield  {author} {\bibinfo {author} {\bibfnamefont {T.~W.~B.}\
  \bibnamefont {Kibble}},\ }\href {\doibase 10.1088/0305-4470/9/8/029}
  {\bibfield  {journal} {\bibinfo  {journal} {J. Phys.}\ }\textbf {\bibinfo
  {volume} {A9}},\ \bibinfo {pages} {1387} (\bibinfo {year}
  {1976})}\BibitemShut {NoStop}%
\bibitem [{\citenamefont {Vilenkin}\ and\ \citenamefont
  {Shellard}(2000)}]{Vilenkin:2000jqa}%
  \BibitemOpen
  \bibfield  {author} {\bibinfo {author} {\bibfnamefont {A.}~\bibnamefont
  {Vilenkin}}\ and\ \bibinfo {author} {\bibfnamefont {E.~P.~S.}\ \bibnamefont
  {Shellard}},\ }\href
  {http://www.cambridge.org/mw/academic/subjects/physics/theoretical-physics-and-mathematical-physics/cosmic-strings-and-other-topological-defects?format=PB}
  {\emph {\bibinfo {title} {{Cosmic Strings and Other Topological Defects}}}}\
  (\bibinfo  {publisher} {Cambridge University Press},\ \bibinfo {year}
  {2000})\BibitemShut {NoStop}%
\bibitem [{\citenamefont {Copeland}\ and\ \citenamefont
  {Kibble}(2010)}]{Copeland:2009ga}%
  \BibitemOpen
  \bibfield  {author} {\bibinfo {author} {\bibfnamefont {E.~J.}\ \bibnamefont
  {Copeland}}\ and\ \bibinfo {author} {\bibfnamefont {T.~W.~B.}\ \bibnamefont
  {Kibble}},\ }\href {\doibase 10.1098/rspa.2009.0591} {\bibfield  {journal}
  {\bibinfo  {journal} {Proc. Roy. Soc. Lond.}\ }\textbf {\bibinfo {volume}
  {A466}},\ \bibinfo {pages} {623} (\bibinfo {year} {2010})},\ \Eprint
  {http://arxiv.org/abs/0911.1345} {arXiv:0911.1345 [hep-th]} \BibitemShut
  {NoStop}%
\bibitem [{\citenamefont {Sakellariadou}(2009)}]{Sakellariadou:2009ev}%
  \BibitemOpen
  \bibfield  {author} {\bibinfo {author} {\bibfnamefont {M.}~\bibnamefont
  {Sakellariadou}},\ }\bibfield  {booktitle} {\emph {\bibinfo {booktitle}
  {{Cargese 2008, theory and particle physics : The LHC perspective and beyond,
  proceedings of the ESF School in High Energy Physics and Astrophysics,
  Cargese Summer School, Cargese, France, 16-28 June 2008}}},\ }\href {\doibase
  10.1016/j.nuclphysbps.2009.07.046} {\bibfield  {journal} {\bibinfo  {journal}
  {Nucl. Phys. Proc. Suppl.}\ }\textbf {\bibinfo {volume} {192-193}},\ \bibinfo
  {pages} {68} (\bibinfo {year} {2009})},\ \Eprint
  {http://arxiv.org/abs/0902.0569} {arXiv:0902.0569 [hep-th]} \BibitemShut
  {NoStop}%
\bibitem [{\citenamefont {Copeland}\ \emph {et~al.}(2011)\citenamefont
  {Copeland}, \citenamefont {Pogosian},\ and\ \citenamefont
  {Vachaspati}}]{Copeland:2011dx}%
  \BibitemOpen
  \bibfield  {author} {\bibinfo {author} {\bibfnamefont {E.~J.}\ \bibnamefont
  {Copeland}}, \bibinfo {author} {\bibfnamefont {L.}~\bibnamefont {Pogosian}},
  \ and\ \bibinfo {author} {\bibfnamefont {T.}~\bibnamefont {Vachaspati}},\
  }\href {\doibase 10.1088/0264-9381/28/20/204009} {\bibfield  {journal}
  {\bibinfo  {journal} {Class. Quant. Grav.}\ }\textbf {\bibinfo {volume}
  {28}},\ \bibinfo {pages} {204009} (\bibinfo {year} {2011})},\ \Eprint
  {http://arxiv.org/abs/1105.0207} {arXiv:1105.0207 [hep-th]} \BibitemShut
  {NoStop}%
\bibitem [{\citenamefont {Vachaspati}\ and\ \citenamefont
  {Vilenkin}(1985)}]{Vachaspati:1984gt}%
  \BibitemOpen
  \bibfield  {author} {\bibinfo {author} {\bibfnamefont {T.}~\bibnamefont
  {Vachaspati}}\ and\ \bibinfo {author} {\bibfnamefont {A.}~\bibnamefont
  {Vilenkin}},\ }\href {\doibase 10.1103/PhysRevD.31.3052} {\bibfield
  {journal} {\bibinfo  {journal} {Phys. Rev.}\ }\textbf {\bibinfo {volume}
  {D31}},\ \bibinfo {pages} {3052} (\bibinfo {year} {1985})}\BibitemShut
  {NoStop}%
\bibitem [{\citenamefont {Hawking}(1989)}]{Hawking:1987bn}%
  \BibitemOpen
  \bibfield  {author} {\bibinfo {author} {\bibfnamefont {S.~W.}\ \bibnamefont
  {Hawking}},\ }\href {\doibase 10.1016/0370-2693(89)90206-2} {\bibfield
  {journal} {\bibinfo  {journal} {Phys. Lett.}\ }\textbf {\bibinfo {volume}
  {B231}},\ \bibinfo {pages} {237} (\bibinfo {year} {1989})}\BibitemShut
  {NoStop}%
\bibitem [{\citenamefont {Polnarev}\ and\ \citenamefont
  {Zembowicz}(1991)}]{Polnarev:1988dh}%
  \BibitemOpen
  \bibfield  {author} {\bibinfo {author} {\bibfnamefont {A.}~\bibnamefont
  {Polnarev}}\ and\ \bibinfo {author} {\bibfnamefont {R.}~\bibnamefont
  {Zembowicz}},\ }\href {\doibase 10.1103/PhysRevD.43.1106} {\bibfield
  {journal} {\bibinfo  {journal} {Phys. Rev.}\ }\textbf {\bibinfo {volume}
  {D43}},\ \bibinfo {pages} {1106} (\bibinfo {year} {1991})}\BibitemShut
  {NoStop}%
\bibitem [{\citenamefont {Carr}\ \emph {et~al.}(2010)\citenamefont {Carr},
  \citenamefont {Kohri}, \citenamefont {Sendouda},\ and\ \citenamefont
  {Yokoyama}}]{Carr:2009jm}%
  \BibitemOpen
  \bibfield  {author} {\bibinfo {author} {\bibfnamefont {B.~J.}\ \bibnamefont
  {Carr}}, \bibinfo {author} {\bibfnamefont {K.}~\bibnamefont {Kohri}},
  \bibinfo {author} {\bibfnamefont {Y.}~\bibnamefont {Sendouda}}, \ and\
  \bibinfo {author} {\bibfnamefont {J.}~\bibnamefont {Yokoyama}},\ }\href
  {\doibase 10.1103/PhysRevD.81.104019} {\bibfield  {journal} {\bibinfo
  {journal} {Phys. Rev.}\ }\textbf {\bibinfo {volume} {D81}},\ \bibinfo {pages}
  {104019} (\bibinfo {year} {2010})},\ \Eprint {http://arxiv.org/abs/0912.5297}
  {arXiv:0912.5297 [astro-ph.CO]} \BibitemShut {NoStop}%
\bibitem [{\citenamefont {Caldwell}\ and\ \citenamefont
  {Gates}(1993)}]{Caldwell:1993kv}%
  \BibitemOpen
  \bibfield  {author} {\bibinfo {author} {\bibfnamefont {R.~R.}\ \bibnamefont
  {Caldwell}}\ and\ \bibinfo {author} {\bibfnamefont {E.}~\bibnamefont
  {Gates}},\ }\href {\doibase 10.1103/PhysRevD.48.2581} {\bibfield  {journal}
  {\bibinfo  {journal} {Phys. Rev.}\ }\textbf {\bibinfo {volume} {D48}},\
  \bibinfo {pages} {2581} (\bibinfo {year} {1993})},\ \Eprint
  {http://arxiv.org/abs/hep-ph/9306221} {arXiv:hep-ph/9306221 [hep-ph]}
  \BibitemShut {NoStop}%
\bibitem [{\citenamefont {MacGibbon}\ \emph {et~al.}(1998)\citenamefont
  {MacGibbon}, \citenamefont {Brandenberger},\ and\ \citenamefont
  {Wichoski}}]{MacGibbon:1997pu}%
  \BibitemOpen
  \bibfield  {author} {\bibinfo {author} {\bibfnamefont {J.~H.}\ \bibnamefont
  {MacGibbon}}, \bibinfo {author} {\bibfnamefont {R.~H.}\ \bibnamefont
  {Brandenberger}}, \ and\ \bibinfo {author} {\bibfnamefont {U.~F.}\
  \bibnamefont {Wichoski}},\ }\href {\doibase 10.1103/PhysRevD.57.2158}
  {\bibfield  {journal} {\bibinfo  {journal} {Phys. Rev.}\ }\textbf {\bibinfo
  {volume} {D57}},\ \bibinfo {pages} {2158} (\bibinfo {year} {1998})},\ \Eprint
  {http://arxiv.org/abs/astro-ph/9707146} {arXiv:astro-ph/9707146 [astro-ph]}
  \BibitemShut {NoStop}%
\bibitem [{\citenamefont {Shannon}\ \emph {et~al.}(2015)\citenamefont {Shannon}
  \emph {et~al.}}]{Shannon:2015ect}%
  \BibitemOpen
  \bibfield  {author} {\bibinfo {author} {\bibfnamefont {R.~M.}\ \bibnamefont
  {Shannon}} \emph {et~al.},\ }\href {\doibase 10.1126/science.aab1910}
  {\bibfield  {journal} {\bibinfo  {journal} {Science}\ }\textbf {\bibinfo
  {volume} {349}},\ \bibinfo {pages} {1522} (\bibinfo {year} {2015})},\ \Eprint
  {http://arxiv.org/abs/1509.07320} {arXiv:1509.07320 [astro-ph.CO]}
  \BibitemShut {NoStop}%
\bibitem [{\citenamefont {Lasky}\ \emph {et~al.}(2016)\citenamefont {Lasky}
  \emph {et~al.}}]{Lasky:2015lej}%
  \BibitemOpen
  \bibfield  {author} {\bibinfo {author} {\bibfnamefont {P.~D.}\ \bibnamefont
  {Lasky}} \emph {et~al.},\ }\href {\doibase 10.1103/PhysRevX.6.011035}
  {\bibfield  {journal} {\bibinfo  {journal} {Phys. Rev.}\ }\textbf {\bibinfo
  {volume} {X6}},\ \bibinfo {pages} {011035} (\bibinfo {year} {2016})},\
  \Eprint {http://arxiv.org/abs/1511.05994} {arXiv:1511.05994 [astro-ph.CO]}
  \BibitemShut {NoStop}%
\bibitem [{\citenamefont {Matzner}(1989)}]{Matzner}%
  \BibitemOpen
  \bibfield  {author} {\bibinfo {author} {\bibfnamefont {R.~A.}\ \bibnamefont
  {Matzner}},\ }\href@noop {} {\bibfield  {journal} {\bibinfo  {journal}
  {Comput. Phys.}\ }\textbf {\bibinfo {volume} {2}},\ \bibinfo {pages} {51}
  (\bibinfo {year} {1989})}\BibitemShut {NoStop}%
\bibitem [{\citenamefont {Blanco-Pillado}\ \emph {et~al.}(2018)\citenamefont
  {Blanco-Pillado}, \citenamefont {Olum},\ and\ \citenamefont
  {Siemens}}]{Blanco-Pillado:2017rnf}%
  \BibitemOpen
  \bibfield  {author} {\bibinfo {author} {\bibfnamefont {J.~J.}\ \bibnamefont
  {Blanco-Pillado}}, \bibinfo {author} {\bibfnamefont {K.~D.}\ \bibnamefont
  {Olum}}, \ and\ \bibinfo {author} {\bibfnamefont {X.}~\bibnamefont
  {Siemens}},\ }\href {\doibase 10.1016/j.physletb.2018.01.050} {\bibfield
  {journal} {\bibinfo  {journal} {Phys. Lett.}\ }\textbf {\bibinfo {volume}
  {B778}},\ \bibinfo {pages} {392} (\bibinfo {year} {2018})},\ \Eprint
  {http://arxiv.org/abs/1709.02434} {arXiv:1709.02434 [astro-ph.CO]}
  \BibitemShut {NoStop}%
\bibitem [{\citenamefont {Caldwell}\ and\ \citenamefont
  {Allen}(1992)}]{Caldwell:1991jj}%
  \BibitemOpen
  \bibfield  {author} {\bibinfo {author} {\bibfnamefont {R.~R.}\ \bibnamefont
  {Caldwell}}\ and\ \bibinfo {author} {\bibfnamefont {B.}~\bibnamefont
  {Allen}},\ }\href {\doibase 10.1103/PhysRevD.45.3447} {\bibfield  {journal}
  {\bibinfo  {journal} {Phys. Rev.}\ }\textbf {\bibinfo {volume} {D45}},\
  \bibinfo {pages} {3447} (\bibinfo {year} {1992})}\BibitemShut {NoStop}%
\bibitem [{\citenamefont {Blanco-Pillado}\ \emph {et~al.}(2014)\citenamefont
  {Blanco-Pillado}, \citenamefont {Olum},\ and\ \citenamefont
  {Shlaer}}]{Blanco-Pillado:2013qja}%
  \BibitemOpen
  \bibfield  {author} {\bibinfo {author} {\bibfnamefont {J.~J.}\ \bibnamefont
  {Blanco-Pillado}}, \bibinfo {author} {\bibfnamefont {K.~D.}\ \bibnamefont
  {Olum}}, \ and\ \bibinfo {author} {\bibfnamefont {B.}~\bibnamefont
  {Shlaer}},\ }\href {\doibase 10.1103/PhysRevD.89.023512} {\bibfield
  {journal} {\bibinfo  {journal} {Phys. Rev. D}\ }\textbf {\bibinfo {volume}
  {89}},\ \bibinfo {pages} {023512} (\bibinfo {year} {2014})},\ \Eprint
  {http://arxiv.org/abs/1309.6637} {arXiv:1309.6637 [astro-ph.CO]} \BibitemShut
  {NoStop}%
\bibitem [{\citenamefont {Blanco-Pillado}\ \emph {et~al.}(2011)\citenamefont
  {Blanco-Pillado}, \citenamefont {Olum},\ and\ \citenamefont
  {Shlaer}}]{BlancoPillado:2011dq}%
  \BibitemOpen
  \bibfield  {author} {\bibinfo {author} {\bibfnamefont {J.~J.}\ \bibnamefont
  {Blanco-Pillado}}, \bibinfo {author} {\bibfnamefont {K.~D.}\ \bibnamefont
  {Olum}}, \ and\ \bibinfo {author} {\bibfnamefont {B.}~\bibnamefont
  {Shlaer}},\ }\href {\doibase 10.1103/PhysRevD.83.083514} {\bibfield
  {journal} {\bibinfo  {journal} {Phys. Rev.}\ }\textbf {\bibinfo {volume}
  {D83}},\ \bibinfo {pages} {083514} (\bibinfo {year} {2011})},\ \Eprint
  {http://arxiv.org/abs/1101.5173} {arXiv:1101.5173 [astro-ph.CO]} \BibitemShut
  {NoStop}%
\bibitem [{\citenamefont {Abbott}\ \emph {et~al.}(2018)\citenamefont {Abbott}
  \emph {et~al.}}]{Abbott:2017mem}%
  \BibitemOpen
  \bibfield  {author} {\bibinfo {author} {\bibfnamefont {B.~P.}\ \bibnamefont
  {Abbott}} \emph {et~al.} (\bibinfo {collaboration} {LIGO Scientific,
  Virgo}),\ }\href {\doibase 10.1103/PhysRevD.97.102002} {\bibfield  {journal}
  {\bibinfo  {journal} {Phys. Rev.}\ }\textbf {\bibinfo {volume} {D97}},\
  \bibinfo {pages} {102002} (\bibinfo {year} {2018})},\ \Eprint
  {http://arxiv.org/abs/1712.01168} {arXiv:1712.01168 [gr-qc]} \BibitemShut
  {NoStop}%
\bibitem [{\citenamefont {Blanco-Pillado}\ \emph {et~al.}(2019)\citenamefont
  {Blanco-Pillado}, \citenamefont {Olum},\ and\ \citenamefont
  {Wachter}}]{Blanco-Pillado:2019vcs}%
  \BibitemOpen
  \bibfield  {author} {\bibinfo {author} {\bibfnamefont {J.~J.}\ \bibnamefont
  {Blanco-Pillado}}, \bibinfo {author} {\bibfnamefont {K.~D.}\ \bibnamefont
  {Olum}}, \ and\ \bibinfo {author} {\bibfnamefont {J.~M.}\ \bibnamefont
  {Wachter}},\ }\href@noop {} {\  (\bibinfo {year} {2019})},\ \Eprint
  {http://arxiv.org/abs/1907.09373} {arXiv:1907.09373 [astro-ph.CO]}
  \BibitemShut {NoStop}%
\bibitem [{\citenamefont {Caldwell}\ and\ \citenamefont
  {Casper}(1996)}]{Caldwell:1995fu}%
  \BibitemOpen
  \bibfield  {author} {\bibinfo {author} {\bibfnamefont {R.~R.}\ \bibnamefont
  {Caldwell}}\ and\ \bibinfo {author} {\bibfnamefont {P.}~\bibnamefont
  {Casper}},\ }\href {\doibase 10.1103/PhysRevD.53.3002} {\bibfield  {journal}
  {\bibinfo  {journal} {Phys. Rev.}\ }\textbf {\bibinfo {volume} {D53}},\
  \bibinfo {pages} {3002} (\bibinfo {year} {1996})},\ \Eprint
  {http://arxiv.org/abs/gr-qc/9509012} {arXiv:gr-qc/9509012 [gr-qc]}
  \BibitemShut {NoStop}%
\bibitem [{\citenamefont {Helfer}\ \emph {et~al.}(2019)\citenamefont {Helfer},
  \citenamefont {Aurrekoetxea},\ and\ \citenamefont {Lim}}]{Helfer:2018qgv}%
  \BibitemOpen
  \bibfield  {author} {\bibinfo {author} {\bibfnamefont {T.}~\bibnamefont
  {Helfer}}, \bibinfo {author} {\bibfnamefont {J.~C.}\ \bibnamefont
  {Aurrekoetxea}}, \ and\ \bibinfo {author} {\bibfnamefont {E.~A.}\
  \bibnamefont {Lim}},\ }\href {\doibase 10.1103/PhysRevD.99.104028} {\bibfield
   {journal} {\bibinfo  {journal} {Phys. Rev.}\ }\textbf {\bibinfo {volume}
  {D99}},\ \bibinfo {pages} {104028} (\bibinfo {year} {2019})},\ \Eprint
  {http://arxiv.org/abs/1808.06678} {arXiv:1808.06678 [gr-qc]} \BibitemShut
  {NoStop}%
\bibitem [{\citenamefont {Carr}\ \emph {et~al.}(2016)\citenamefont {Carr},
  \citenamefont {Kohri}, \citenamefont {Sendouda},\ and\ \citenamefont
  {Yokoyama}}]{Carr:2016hva}%
  \BibitemOpen
  \bibfield  {author} {\bibinfo {author} {\bibfnamefont {B.~J.}\ \bibnamefont
  {Carr}}, \bibinfo {author} {\bibfnamefont {K.}~\bibnamefont {Kohri}},
  \bibinfo {author} {\bibfnamefont {Y.}~\bibnamefont {Sendouda}}, \ and\
  \bibinfo {author} {\bibfnamefont {J.}~\bibnamefont {Yokoyama}},\ }\href
  {\doibase 10.1103/PhysRevD.94.044029} {\bibfield  {journal} {\bibinfo
  {journal} {Phys. Rev.}\ }\textbf {\bibinfo {volume} {D94}},\ \bibinfo {pages}
  {044029} (\bibinfo {year} {2016})},\ \Eprint
  {http://arxiv.org/abs/1604.05349} {arXiv:1604.05349 [astro-ph.CO]}
  \BibitemShut {NoStop}%
\bibitem [{\citenamefont {MacGibbon}\ \emph {et~al.}(2008)\citenamefont
  {MacGibbon}, \citenamefont {Carr},\ and\ \citenamefont
  {Page}}]{MacGibbon:2007yq}%
  \BibitemOpen
  \bibfield  {author} {\bibinfo {author} {\bibfnamefont {J.~H.}\ \bibnamefont
  {MacGibbon}}, \bibinfo {author} {\bibfnamefont {B.~J.}\ \bibnamefont {Carr}},
  \ and\ \bibinfo {author} {\bibfnamefont {D.~N.}\ \bibnamefont {Page}},\
  }\href {\doibase 10.1103/PhysRevD.78.064043} {\bibfield  {journal} {\bibinfo
  {journal} {Phys. Rev.}\ }\textbf {\bibinfo {volume} {D78}},\ \bibinfo {pages}
  {064043} (\bibinfo {year} {2008})},\ \Eprint {http://arxiv.org/abs/0709.2380}
  {arXiv:0709.2380 [astro-ph]} \BibitemShut {NoStop}%
\bibitem [{\citenamefont {Kolb}\ and\ \citenamefont
  {Turner}(1990)}]{Kolb:1990vq}%
  \BibitemOpen
  \bibfield  {author} {\bibinfo {author} {\bibfnamefont {E.~W.}\ \bibnamefont
  {Kolb}}\ and\ \bibinfo {author} {\bibfnamefont {M.~S.}\ \bibnamefont
  {Turner}},\ }\href@noop {} {\bibfield  {journal} {\bibinfo  {journal} {Front.
  Phys.}\ }\textbf {\bibinfo {volume} {69}},\ \bibinfo {pages} {1} (\bibinfo
  {year} {1990})}\BibitemShut {NoStop}%
\bibitem [{\citenamefont {Aghanim}\ \emph {et~al.}(2018)\citenamefont {Aghanim}
  \emph {et~al.}}]{Aghanim:2018eyx}%
  \BibitemOpen
  \bibfield  {author} {\bibinfo {author} {\bibfnamefont {N.}~\bibnamefont
  {Aghanim}} \emph {et~al.} (\bibinfo {collaboration} {Planck}),\ }\href@noop
  {} {\  (\bibinfo {year} {2018})},\ \Eprint {http://arxiv.org/abs/1807.06209}
  {arXiv:1807.06209 [astro-ph.CO]} \BibitemShut {NoStop}%
\bibitem [{\citenamefont {Barrau}\ \emph {et~al.}(2003)\citenamefont {Barrau},
  \citenamefont {Boudoul},\ and\ \citenamefont {Derome}}]{Barrau:2003nj}%
  \BibitemOpen
  \bibfield  {author} {\bibinfo {author} {\bibfnamefont {A.}~\bibnamefont
  {Barrau}}, \bibinfo {author} {\bibfnamefont {G.}~\bibnamefont {Boudoul}}, \
  and\ \bibinfo {author} {\bibfnamefont {L.}~\bibnamefont {Derome}},\ }in\
  \href {http://www-rccn.icrr.u-tokyo.ac.jp/icrc2003/PROCEEDINGS/PDF/421.pdf}
  {\emph {\bibinfo {booktitle} {{Proceedings, 28th International Cosmic Ray
  Conference (ICRC 2003): Tsukuba, Japan, July 31-August 7, 2003}}}}\ (\bibinfo
  {year} {2003})\ pp.\ \bibinfo {pages} {1697--1699},\ \Eprint
  {http://arxiv.org/abs/astro-ph/0304528} {arXiv:astro-ph/0304528 [astro-ph]}
  \BibitemShut {NoStop}%
\bibitem [{\citenamefont {Ballesteros}\ \emph {et~al.}(2019)\citenamefont
  {Ballesteros}, \citenamefont {Coronado-Blázquez},\ and\ \citenamefont
  {Gaggero}}]{Ballesteros:2019exr}%
  \BibitemOpen
  \bibfield  {author} {\bibinfo {author} {\bibfnamefont {G.}~\bibnamefont
  {Ballesteros}}, \bibinfo {author} {\bibfnamefont {J.}~\bibnamefont
  {Coronado-Blázquez}}, \ and\ \bibinfo {author} {\bibfnamefont
  {D.}~\bibnamefont {Gaggero}},\ }\href@noop {} {\  (\bibinfo {year} {2019})},\
  \Eprint {http://arxiv.org/abs/1906.10113} {arXiv:1906.10113 [astro-ph.CO]}
  \BibitemShut {NoStop}%
\bibitem [{\citenamefont {Carr}\ \emph {et~al.}(2017)\citenamefont {Carr},
  \citenamefont {Raidal}, \citenamefont {Tenkanen}, \citenamefont {Vaskonen},\
  and\ \citenamefont {Veermäe}}]{Carr:2017jsz}%
  \BibitemOpen
  \bibfield  {author} {\bibinfo {author} {\bibfnamefont {B.}~\bibnamefont
  {Carr}}, \bibinfo {author} {\bibfnamefont {M.}~\bibnamefont {Raidal}},
  \bibinfo {author} {\bibfnamefont {T.}~\bibnamefont {Tenkanen}}, \bibinfo
  {author} {\bibfnamefont {V.}~\bibnamefont {Vaskonen}}, \ and\ \bibinfo
  {author} {\bibfnamefont {H.}~\bibnamefont {Veermäe}},\ }\href {\doibase
  10.1103/PhysRevD.96.023514} {\bibfield  {journal} {\bibinfo  {journal} {Phys.
  Rev.}\ }\textbf {\bibinfo {volume} {D96}},\ \bibinfo {pages} {023514}
  (\bibinfo {year} {2017})},\ \Eprint {http://arxiv.org/abs/1705.05567}
  {arXiv:1705.05567 [astro-ph.CO]} \BibitemShut {NoStop}%
\end{thebibliography}%

\end{document}